\def\p{{\bf p}}
\def\q{{\bf q}}
\def\F{{\bf F}}
\def\E{{\bf E}}
\def\Fito{\tilde F}
\def\bFito{\tilde\F}
\def\bxi{\mbox{\boldmath$\xi$}}
\def\tr{{\rm tr}}
\def\W{^{}_{\rm W}}
\def\P{{\cal P}}

\def\half{\textstyle{1\over2}}

\def\eighth{\textstyle{1\over8}}

\documentstyle[preprint,aps,eqsecnum,fixes,epsfig,amstex]{revtex}
\tightenlines
\begin {document}


\preprint {UVA/Arnold--99--43}

\title
{Symmetric path integrals for stochastic equations with multiplicative noise}

\author {Peter Arnold}

\address
    {%
    Department of Physics,
    University of Virginia,
    Charlottesville, VA 22901
    }%
\date {November 1999}

\maketitle
\vskip -20pt

\begin {abstract}%
{%
A Langevin equation with multiplicative noise is an equation schematically
of the form ${d\q/dt} = - \F(\q) + e(\q)\, \bxi$,
where $e(\q)\, \bxi$ is Gaussian
white noise whose amplitude $e(\q)$ depends on $\q$ itself.
I show how to convert such equations into path integrals.
The definition of the path integral depends crucially
on the convention used for discretizing time, and I specifically
derive the correct path integral when the convention used is the
natural, time-symmetric one that
time derivatives are $(\q_t - \q_{t-\Delta t})/\Delta t$ and coordinates
are $(\q_t + \q_{t-\Delta t})/2$.
[This is the convention that permits standard manipulations of calculus
on the action, like naive integration by parts.]
It has sometimes been assumed in the literature that a Stratanovich
Langevin equation can be quickly converted to a path integral by
treating time as continuous but using the rule $\theta(t{=}0) = {1\over2}$.
I show that this prescription fails when the amplitude $e(\q)$ is
$\q$-dependent.
}%
\end {abstract}

\thispagestyle{empty}


\section {Introduction}

Let $\bxi$ be Gaussian white noise, which I'll normalize as
\begin {equation}
   \langle \xi_i(t) \, \xi_j(t') \rangle = \Omega \, \delta(t-t') .
\label {eq:xi}
\end {equation}
It's long been known that a Langevin equation of the form
\begin {equation}
   {d\over dt} \, q_i = - F_i(\q) + \xi_i
\end {equation}
can be alternatively described in terms of a path integral of the form%
\footnote{
   For a review of background material in notation close to that I use here,
   see, for
   example, chapter 4 of ref.\ \cite{Zinn-Justin}.  The most substantial
   difference in notation is that my $\F$ is that reference's
   ${1\over 2}{\bf f}$.
}
\begin {equation}
   \P(\q'',\q',t''-t')
   =  \int_{\q(t')=\q'}^{\q(t'')=\q''}
      \left[d\q(t)\right] \>
      \exp\left[ - \int_{t'}^{t''} L(\dot\q,\q) \right] ,
\end {equation}
Here, $\P(\q'',\q',t)$ is the probability density that the system will end up
at $\q''$ at time $t$ if it started at $\q'$ at time zero.
However, the exact form of $L$ depends on the convention used in
discretizing time when defining the path integral.
With a symmetric discretization,%
\footnote{
  For a discussion of what changes if other discretizations are used in this
  case, see refs.\ \cite{other}
}
\begin {equation}
   \P(\q'',\q',t)
   = \lim_{\Delta t \to 0} N
      \int_{\q(0)=\q'}^{\q(t)=\q''}
      \left[\prod_t d\q_t \right] \>
      \exp\left[ - \Delta t \sum_t
          L\left( {\q_t - \q_{t-\Delta t} \over \Delta t} \, , \,
                  {\q_t + \q_{t-\Delta t} \over 2} \right)
      \right] ,
\label{eq:path1}
\end {equation}
with Lagrangian
\begin {equation}
   L(\dot\q, \q) = {1\over 2\Omega} \, |\dot\q + \F|^2 - {1\over2}\, F_{i,i} .
\end {equation}
Here and throughout, I adopt the notation that indices after a comma
represent derivatives: $F_{i,j} \equiv \partial F_i/\partial q_j$
and $F_{i,jk} \equiv \partial^2 F_i/\partial q_j \partial q_k$.
$N$ is the usual overall normalization of the path integral,
which I will not bother being explicit about.

What hasn't been properly discussed, to my knowledge, is how to correctly
form such a symmetrically-discretized path integral for the case of
Langevin equations with multiplicative noise (meaning noise whose amplitude
$e(\q)$ depends on $\q$).  Schematically,
\begin {equation}
   {d\over dt} \, q_i = - F_i(\q) + e_{ia}(\q) \, \xi_a ,
\label{eq:Langevin}
\end {equation}
with $\xi$ as before (\ref{eq:xi}).
I will assume that the matrix $e_{ia}$ is invertible.
There are a wide variety of applications of such equations, but I'll just
mention one particular example of interest to me, which motivated this
work and for which a path integral formulation is particularly convenient:
the calculation of the rate of electroweak baryon number violation in the
early universe \cite{arnold}.

By itself, the continuum equation (\ref{eq:Langevin}) suffers a well-known
ambiguity.
To define the problem more clearly, we must discretize time and take
$\Delta t \to 0$ at the end of the day.  Specifically, I will interpret
(\ref{eq:Langevin}) in Stratanovich convention, writing
\begin {equation}
   \q_t - \q_{t-\Delta t}
   = - \Delta t \, \F(\bar\q) + e(\bar\q) \, \bxi_t ,
\label {eq:LangStrat}
\end {equation}
\begin {equation}
   \langle \xi_{at} \xi_{bt'} \rangle
   = \Delta t\, \Omega \, \delta_{ab} \delta_{t t'} ,
\end {equation}
where
\begin {equation}
   \bar\q \equiv {\q_t+\q_{t-\Delta t} \over 2} \,.
\end {equation}
[The fact that I've labelled the noise $\bxi_t$ instead of
$\bxi_{t-\Delta t}$ in (\ref{eq:LangStrat})
is just an inessential choice of convention.]
The Stratanovich equation (\ref{eq:LangStrat})
is equivalent to the It\^o equation
\begin {equation}
   \q_t - \q_{t-\Delta t}
   = - \Delta t \, \bFito_i(\q_{t-\Delta t}) + e(\q_{t-\Delta t}) \, \bxi_t ,
\label {eq:LangIto}
\end {equation}
with
\begin {equation}
   \Fito_i = F_i - {\Omega\over2} \, e_{ia,j} e_{ja} .
\end {equation}

I will give two different methods for deriving the corresponding
path integral.
The result is
\begin {eqnarray}
   \P(\q'',\q',t)
   &=& \lim_{\Delta t \to 0} N
      \int_{\q(0)=\q'}^{\q(t)=\q''}
      \left[\prod_t d\q_t \right] \>
      \left[\prod_t \det e\!\left(\q_t + \q_{t-\Delta t} \over 2\right)
            \right]^{-1}
\nonumber\\ && \qquad\qquad \times
      \exp\left[ - \Delta t \sum_t
          L\left( {\q_t - \q_{t-\Delta t} \over \Delta t} \, , \,
                  {\q_t + \q_{t-\Delta t} \over 2} \right)
      \right] ,
\label{eq:path2}
\end {eqnarray}
\begin {equation}
   L(\dot\q, \q) = {1\over 2\Omega} \, (\dot q + F)_i g_{ij} (\dot q + F)_j
      - {1\over2} \, F_{i,i}
      + {1\over2} \, e^{-1}_{ia} e_{ka,k}(\dot q + F)_i
      + {\Omega\over8} \, e_{ia,j} e_{ja,i} \, ,
\label{eq:Lresult}
\end {equation}
where
\begin {equation}
   g_{ij} \equiv (e^{-1})_{ia} (e^{-1})_{ja} .
\end {equation}
This differs from  a result previously given by
Zinn-Justin%
\footnote{
  Specifically, eq.\ (4.79) or ref.\ \cite{Zinn-Justin}.
  See also ref.\ \cite{Zinn-Justin1} and section 17.8 or
  ref.\ \cite{Zinn-Justin} for a continuum time discussion of formulating
  the path integral for this problem using ghosts.
}
\cite{Zinn-Justin}
by the inclusion of the last term in $L$.
Zinn-Justin's derivation was done in continuous time, resolving ambiguities
using the prescription $\theta(t{=}0) = {1\over2}$, which is known to work
in the case where $e(\q)$ is constant.

Because of the confusion surrounding these issues, I will show how to
derive the result in two different ways.  First, I will follow the
standard procedure for directly turning Langevin equations into
path integrals, but I will be careful to keep time discrete throughout
the derivation.  The second method will be to start from the
Fokker-Planck equation equivalent to the Langevin equations
(\ref{eq:LangStrat}) and (\ref{eq:LangIto})
and to then turn that Fokker-Planck equation
into a path integral, again  using standard methods.


\section{Direct derivation from the Langevin equation}

Rewrite the discretized Langevin equation (\ref{eq:LangStrat}) as
\begin {equation}
   \E_\tau - \bxi_\tau = 0 ,
\end {equation}
where $\tau$ is a discrete time index and
\begin {equation}
   \E_{\tau} \equiv 
   e^{-1}(\bar\q) \, [\q_\tau - \q_{\tau-1} + \Delta t\, \F(\bar\q)] ,
\label {eq:Edef}
\end {equation}
with $\bar\q \equiv (\q_\tau + \q_{\tau-1})/2$.
The corresponding path integral is obtained by implementing these equations,
for each
value of $\tau$, as $\delta$-functions, with appropriate Jacobian,
integrated over the Gaussian noise distribution:
\begin {equation}
   \P(\q'',\q',t)
   = \lim_{\Delta t \to 0} N
    \int_{\q(0)=\q'}^{\q(t)=\q''}
    \left[ \prod_\tau
      d\bxi_\tau \> \exp\!\left(-{\xi_\tau^2\over 2\Omega\,\Delta t} \right) \>
      d\q_\tau \> \delta(E_\tau - \xi_\tau) 
    \right]
    \det_{\tau' a; \tau'' i}\!
        \left( \partial E_{\tau' a} \over \partial q_{\tau'' i} \right)
   .
\end {equation}
The noise integral then gives
\begin {eqnarray}
   \P(\q'',\q',t)
   =
   \int \left[ \prod_\tau d\q_\tau \right]
        \exp\!\left( - {1\over 2\Omega\,\Delta t} \sum_\tau E_\tau^2 \right) \,
        \det_{\tau' a;\tau'' i}
           \left(\partial E_{\tau' a} \over \partial q_{\tau'' i}\right)
   .
\end {eqnarray}
In our case (\ref{eq:Edef}), the determinant takes the form
\begin {equation}
    \det_{\tau'a;\tau''i}
       \left(\partial E_{\tau' a} \over \partial q_{\tau'' i}\right) =
    \det
    \begin{pmatrix}
       {\partial\E_1 \over \partial\q_1}
             & 0 & 0 & 0 & \\[3pt]
       {\partial\E_2 \over \partial\q_1}
             & {\partial\E_2 \over \partial\q_2}
             & 0 & 0 & \\[3pt]
       0 & {\partial\E_3 \over \partial\q_2}
             & {\partial\E_3 \over \partial\q_3}
             & 0 & \\[3pt]
       0 & 0 & {\partial\E_4 \over \partial\q_3}
             & {\partial\E_4 \over \partial\q_4}
             & \\[3pt]
       & & & \ddots & \ddots
   \end{pmatrix}
   = \prod_\tau \det_{ai} \!
         \left({\partial E_{\tau a} \over \partial q_{\tau i}}\right)
   .
\end {equation}
The registration of the diagonals is determined by the nature of the initial
boundary condition, which is that $q_0$ is fixed.
From (\ref{eq:Edef}), we then have
\begin {eqnarray}
&&
    \det_{\tau' a;\tau'' i}
       \left(\partial E_{\tau' a} \over \partial q_{\tau'' i}\right) =
    \prod_\tau \det_{ai} \left[ (e^{-1})_{ai}
                      + \half \, (e^{-1})_{ak,i}
                             \, (q_\tau - q_{\tau-1})_k
                      + \Delta t \, \half (e^{-1}\F)_{a,i} \right]
\nonumber\\ && \quad
    =
    \left\{\prod_\tau \det\left[e^{-1}(\bar\q_\tau)\right] \right\}
    \left\{\prod_\tau \det_{ij} \left[ \delta_{ij}
                      + \half \, e_{ja} (e^{-1})_{ak,i}
                             \, (q_\tau - q_{\tau-1})_k
                      + \Delta t \, \half e_{ja} (e^{-1}\F)_{a,i}
           \right] \right\} ,
\label {eq:blup}
\end {eqnarray}
where all $e$'s and $F$'s should now be understood as evaluated at
$\bar\q_\tau$.
Now rewrite the determinants in the last factor of (\ref{eq:blup})
as exponentials in the usual way, using
\begin {equation}
   \det(1+A) = e^{\tr\ln(1+A)}
   = \exp\,\tr\left[1 + A - \half A^2 + \cdots \right] .
\label{eq:expand}
\end {equation}
To construct a path integral, we need to keep track of the terms in each
time step up to and including $O(\Delta t)$, but we can ignore corrections
that are higher-order in $\Delta t$.  For this purpose, the size of
$\q_\tau - \q_{\tau-1}$ should be treated as $O(\sqrt{\Delta t})$, which
is the size for which the $\dot q^2$ term in the action [the exponent in
(\ref{eq:path2})] becomes $O(1)$ per degree of freedom.
So, using the expansion (\ref{eq:expand}), we get
\begin {eqnarray}
&&
\det_{ij} \left[ \delta_{ij}
                      + \half \, e_{ja} (e^{-1})_{ak,i}
                             \, \Delta q_k
                      + \Delta t \, \half e_{ja} (e^{-1}\F)_{a,i}
           \right]
\nonumber\\ && \qquad
    = \exp\Bigl\{ \mbox{(constant)} + 
                        \half \, e_{ia} (e^{-1})_{ak,i}
                             \Delta q_k
                      + \Delta t \, \half e_{ia} (e^{-1} \F)_{a,i}
\nonumber\\ && \hspace{8em}
                      - \eighth \, e_{ia} (e^{-1})_{ak,m}
                             e_{mb} (e^{-1})_{bl,i}
                             \Delta q_k \,
                             \Delta q_l
                      + O[(\Delta t)^{3/2}] \Bigr\} .
\end {eqnarray}
It is the $\Delta q \, \Delta q$ term in this equation, which came from
the second-order term in the expansion (\ref{eq:expand}), that will
generate the difference with the result quoted in ref.\ \cite{Zinn-Justin}.
Putting everything together, we get the path integral (\ref{eq:path2})
with Lagrangian
\begin {eqnarray}
   L(\dot\q, \q) &=& {1\over 2\Omega} \, (\dot q + F)_i g_{ij} (\dot q + F)_j
      - {1\over2} \, F_{i,i}
      + {1\over2} \, e^{-1}_{ia} e_{ka,k}(\dot q + F)_i
\nonumber\\ && \hspace{8em}
      + {\Delta t\over8} \, e_{ia} (e^{-1})_{ak,m}
                             e_{mb} (e^{-1})_{bl,i}
                             \dot q_k
                             \dot q_l
   .
\label {eq:Lalmost}
\end {eqnarray}
We can simplify this by realizing that the $\dot q_k \dot q_l$ in the
last term can be replaced by its leading-order behavior in $\Delta t$.
Specifically, recall that $\q_t-\q_{t-\Delta t}$ is order $\sqrt{\Delta t}$.
So one can go for a large number of discrete time
steps $1 \ll N \ll 1/\Delta t$
without any net change in $\q$ at leading order in $\Delta t$.
Moreover, the force $\F$ doesn't have any net effect, at leading order in
$\Delta t$, over that number of steps.  The upshot is that $\dot q_k \dot q_l$
can be replaced at leading order in $\Delta t$ by its average over a large
number of steps, ignoring $\F$ and treating the background value of $e(\bar\q)$
as constant.  The Gaussian integral for $\dot\q$ in (\ref{eq:path2}) and
(\ref{eq:Lalmost}) then gives the replacement rule
\begin {equation}
   \dot q_k \dot q_l \to {\Omega\over\Delta t} \, (g^{-1})_{kl}
\end {equation}
at leading order in $\Delta t$.
This substitution turns (\ref{eq:Lalmost}) into the result (\ref{eq:Lresult})
presented earlier.


\section {Derivation from Fokker-Planck equation}

The Stratanovich Langevin equation (\ref{eq:LangStrat})
is well-known to be equivalent to the Fokker-Planck
equation
\begin {equation}
   \dot P = {\partial\over\partial q_i} \left[
      {\Omega\over2}\, e_{ia} {\partial\over\partial q_j} (e_{ja} P)
                                     + F_i P \right] .
\end {equation}
where $P = P(\q,t)$ is the probability distribution of the system as a
function of time.
This is just a Euclidean Schr\"odinger equation, and one can transform
Schr\"odinger equations into path integrals by standard methods.
Specifically, rewrite the equation as
\begin {equation}
   \dot P = - \hat H P ,
\end {equation}
with the Hamiltonian
\begin {equation}
   \hat H = {\Omega\over2}\, \hat p_i \, e_{ia}(\hat\q) \, \hat p_j
                                 \, e_{ja}(\hat\q)
                         - i \hat\p\cdot\F(\hat\q) .
\label {eq:H0}
\end {equation}

To obtain a path integral with symmetric time discretization, it is
be convenient to rewrite $\hat H$ in terms of Weyl-ordered operators.
The Weyl order corresponding to a classical expression
${\cal O}(\p,\q,t)$ is defined
as the operator $\hat{\cal O}\W$ with
\begin {equation}
   \langle q | \hat{\cal O}\W | q' \rangle
   = \int_\p e^{i\p\cdot(\q'-\q)} \, O\!\left(\p,\half(\q+\q'),t\right) .
\end {equation}
For the sake of completeness, I'll briefly review how to obtain
Weyl ordering of operators in simple cases by considering
the application of the operators to an arbitrary function $\psi(\q)$.
For example,
\begin {eqnarray}
   [p_i \, A(\q)]\W \, \psi(\q)
   &=& \langle \q | [p_i \, A(\q)]\W | \psi \rangle
\nonumber\\
   &=& \int_{\q'}  \langle \q | [p_i \, A(\q)]\W
                          | \q' \rangle \, \psi(\q')
\nonumber\\
   &=& \int_{\q'}  \int_\p e^{i\p\cdot(\q-\q')}
              p_i \, A\!\left(\q+\q'\over2\right)
              \psi(\q')
\nonumber\\
   &=& i \int_{\q'}  \left[{\partial\over\partial q'_i} \delta(\q-\q')\right]
              A\!\left(\q+\q'\over2\right)
              \psi(\q')
\nonumber\\
   &=& -i {\partial\over\partial q'_i} \,
              A\!\left(\q+\q'\over2\right)
              \psi(\q') \bigg|_{\q'=\q}
\nonumber\\
   &=& -i \left\{
          {1\over2} \left[ {\partial\over\partial q_i} \, A(\q) \right]
          + A(\q) \, {\partial\over\partial q_i} \right\} \psi(\q) .
\end {eqnarray}
So
\begin {equation}
   [p_i \, A(\q)]\W
   = {1\over2} \left\{ \hat p_i , A(\hat\q) \right\} .
\end {equation}
One can similarly show that
\begin {equation}
   [p_i p_j \, A(\q)]\W
   = {1\over4} \left\{ \hat p_i , \left\{ \hat p_j , A(\hat\q) \right\}
                  \right\}.
\end {equation}

Now write the Hamiltonian (\ref{eq:H0}) in terms of
Weyl-ordered operators.  One finds
\begin {equation}
   \hat\p\cdot\F = [\p\cdot\F]\W - {i\over2}\, F_{i,i} ,
\end {equation}
\begin {equation}
   \hat p_i e_{ia} \hat p_j e_{ja}
   = [\p g^{-1}\p]\W
          - i [e_{ia} e_{ja,j} \hat p_i]\W
          + {1\over4} \, (g^{-1})_{ij,ij}
          - {1\over2} \, (e_{ia} e_{ja,j})_{,i} .
\end {equation}
So
\begin {equation}
   \hat H = [H(\p,\q)]\W ,
\end {equation}
with
\begin {equation}
   H(\p,\q) = {\Omega\over2} \, \p\,g^{-1}(\q)\,\p
              - i \p_i\left[F_i(\q) 
                      + {\Omega\over2} \, e_{ia}(\q)\, e_{ja,j}(\q)\right]
              + u(\q) ,
\label{eq:H1}
\end {equation}
\begin {equation}
   u = - {1\over2}\,F_{i,i} + {\Omega\over8}(g^{-1})_{ij,ij}
          - {\Omega\over4} (e_{ia} e_{ja,j})_{,i} \, .
\end {equation}
The usual derivation of the path integral then gives
\begin {equation}
   \P(\q'',\q',t) =
   \lim_{\Delta t \to 0} \int_{\q(0)=\q'}^{\q(t)=\q''}
     \left[\prod_\tau {d \p_\tau \> d\q_\tau \over (2\pi)^d}\right]
     e^{-S(\p,\q)} ,
\end {equation}
\begin {equation}
   S(\p,\q) = \sum_\tau \left\{
             -i \p_\tau \cdot (\q_\tau - \q_{\tau-1})
              + \Delta t \,
              H\!\left(\p_\tau,{\q_\tau + \q_{\tau-1}\over2}\right)
             \right\} ,
\end {equation}
Doing the $\p$ integrals with our Hamiltonian (\ref{eq:H1})
then reproduces the path integral (\ref{eq:path1}) with Lagrangian
\begin {equation}
   L(\dot\q, \q) =
        {1\over 2\Omega} \,
           \left( \dot q_i + F_i + \half \Omega e_{ia} e_{ka,k} \right) 
           g_{ij}
           \left( \dot q_j + F_j + \half \Omega e_{jb} e_{lb,l} \right)
        + u(\q) .
\label {eq:last1}
\end {equation}
Now note that
\begin {equation}
   (g^{-1})_{ij,ij} = 2 e_{ia} e_{ja,ij} + e_{ia,i} e_{ja,j}
            + e_{ia,j} e_{ja,i} ,
\end {equation}
and so
\begin {eqnarray}
   {\Omega\over8} (e_{ia} e_{ka,k}) g_{ij} (e_{jb} e_{lb,l})
          + u
   &=& {\Omega\over8} \, e_{ia,i} e_{ja,j}
      + \left[
        - {1\over2} \, F_{i,i}
        + {\Omega\over8} \, (g^{-1})_{ij,ij}
        - {\Omega\over4} \, (e_{ia} e_{ja,j})_{,i}
     \right]
\nonumber\\
   &=&  - {1\over2} \, F_{i,i}
      + {\Omega\over8} \, e_{ia,j} e_{ja,i} .
\label {eq:last2}
\end {eqnarray}
Combining (\ref{eq:last1}) and (\ref{eq:last2}) reproduces
the Lagrangian (\ref{eq:Lresult}) asserted in the introduction.


\section* {ACKNOWLEDGMENTS}

I thank Larry Yaffe, Dam Son, and Tim Newman for useful conversations.
This work was supported by the U.S. Department
of Energy under Grant Nos.~DEFG02-97ER41027.


\end {document}